\documentstyle[prl,aps,preprint]{revtex}
\tightenlines
\begin{document}
\newcommand{\beq}{\begin{equation}}
\newcommand{\eeq}{\end{equation}}
\newcommand{\barr}{\begin{eqnarray}}
\newcommand{\earr}{\end{eqnarray}}

\newcommand{\andy}[1]{ }

\def\txt{\textstyle}
\def\eqn#1{Eq.\ (\ref{eq:#1})} \def\pp{{\vec p}}
\def\ask{\marginpar{?? ask:  \hfill}}  
\def\fin{\marginpar{fill in ... \hfill}}
\def\note{\marginpar{note \hfill}}  
\def\check{\marginpar{check \hfill}}
\def\discuss{\marginpar{discuss \hfill}}
\def\hh{\widehat}
\def\tt{\widetilde}
\def\cH{{\cal H}}
\def\cT{{\cal T}}
\def\cR{{\cal R}}
\def\cC{{\cal C}}
\def\cZ{{\cal Z}}
\def\cL{{\cal L}}
\def\txt{\textstyle}
\def\bmp{\mbox{\boldmath $p$}}
\def\bmk{\mbox{\boldmath $k$}}
\def\bmksub{\mbox{\boldmath\scriptsize $k$}}
\def\bmp{\mbox{\boldmath $p$}}
\def\bmr{\mbox{\boldmath $r$}}
\def\bmA{\mbox{\boldmath $A$}}
\def\bmv{\mbox{\boldmath $v$}}
\def\bmg{\mbox{\boldmath $g$}}
\def\bmepsilon{\mbox{\boldmath $\epsilon$}}
\def\bmhA{\mbox{\boldmath $\hat{A}$}}
\def\bmhp{\mbox{\boldmath $\hat{p}$}}
\def\bmhv{\mbox{\boldmath $\hat{v}$}}
\def\pade#1#2{{frac{\partial#1}{\partial#2}}}
\newcommand{\mean}[1]{\langle #1 \rangle}


\draft

\title{ Measurement-induced quantum diffusion}
\author{ P. FACCHI,$^{1,2}$ 
         S. PASCAZIO$^{1,2}$ 
         and A. SCARDICCHIO$^{1}$
         }
\address{$^{1}$Dipartimento di Fisica, Universit\`a di Bari 
I-70126  Bari, Italy \\ $^{2}$Istituto Nazionale di Fisica 
Nucleare, Sezione di Bari, I-70126  Bari, Italy} 

\date{\today}

\maketitle

\begin{abstract} 
The dynamics of a ``kicked" quantum system undergoing repeated
measurements of momentum is investigated. A diffusive behavior is 
obtained even when the dynamics of the classical counterpart is not 
chaotic. The diffusion coefficient is explicitly computed for a 
large class of Hamiltonians and compared to the classical case.
\end{abstract}


The classical and quantum dynamics of bound Hamiltonian systems
under the action of periodic ``kicks" are in general very 
different. Classical systems can follow very complicated 
trajectories in phase space, while the evolution of the wave 
function in the quantum case is more regular. This phenomenon, 
discovered two decades ago \cite{Chirikov,Berry}, as well as the 
features of the quantum mechanical suppression of classical chaos 
and the semiclassical approximation ($\hbar \to 0$) are now well 
understood \cite{CCSG,Tabor}. 

The ``kicked" rotator (standard map) has played an important role in the
study of the different features of the classical and quantum case. 
This model is very useful not only because it elucidates
several conceptual differences between these two cases, 
but also for illustrative purposes. One of the 
most distinctive features of an underlying chaotic behavior is the diffusive
character of the dynamics of the classical action variable in phase space. 
In the quantum case, this diffusion is always suppressed after a sufficiently
long time.
On the other hand, it has been shown \cite{KaulGontis} that, 
in the case of the kicked rotator, a diffusive behavior
is obtained even in the quantum
case, if a measurement is performed after every kick. 
The purpose of this Letter is to investigate this situation in more detail,
focussing our attention on the role played by the measurement process in the
quantum dynamics.
We will see that 
quantum  measurements provoke diffusion in a very large class of ``kicked" 
systems, even when the corresponding classical dynamics is regular.

We consider the following Hamiltonian (in action-angle variables)
\andy{eq:hamilt}
\beq
H=H_0(p) + \lambda V(x) \delta_T(t),
\label{eq:hamilt}
\eeq
where 
\andy{eq:deltat}
\beq
\delta_T(t)=\sum_{k=-\infty}^{\infty} \delta (t-kT),
\label{eq:deltat}
\eeq
$T$ being the period of the perturbation. 
The interaction $V(x)$ is defined for $x \in [-\pi,\pi]$, with 
periodic boundary conditions. This Hamiltonian gives rise to 
the radial twisting map. This is a wide class of maps, 
including as a particular case the standard map \cite{Lichten}, 
which describes the local behavior of a perturbed integrable map
near resonance.
The free Hamiltonian $H_0$ has a discrete spectrum and a countable complete 
set of eigenstates $\{|m\rangle\}$:
\andy{eigenfun}
\beq
\langle x|m\rangle=\frac{1}{\sqrt{2\pi}} \exp\left(imx\right),
\qquad m=0,\pm1,\pm2,\cdots.
\label{eq:eigenfun}
\eeq
We shall consider the evolution engendered by
(\ref{eq:hamilt}) interspersed with quantum
measurements, in the following sense: the system evolves under the action
of the free Hamiltonian for $(N-1)T+\tau<t< NT \;$     ($0<\tau<T$),
undergoes a ``kick" at 
$t=NT$, evolves again freely and
then undergoes a ``measurement" of $p$ at $t=NT+\tau$. 
The evolution of the density matrix between measurements is
\andy{eq:propag1,2}
\barr
\rho_{NT+\tau}&=&U_{\rm free}(\tau) U_{\rm kick} U_{\rm free}(T-\tau)
\rho_{(N-1)T+\tau}U_{\rm free}^\dagger(T-\tau)U_{\rm kick}^\dagger 
U_{\rm free}^\dagger(\tau),
 \label{eq:propag1} \\
& & U_{\rm kick} = \exp\left(-i \lambda V/ \hbar \right), \quad 
U_{\rm free} (t) = \exp\left(-i H_0 t/ \hbar\right).
\label{eq:propag2}
\earr 
At each measurement, the wave function is ``projected" onto 
the $n$th eigenstate of $p$ with probability 
\andy{projproc}
\beq
P_n(NT+\tau)={\rm Tr}(|n\rangle\langle n|\rho_{NT+\tau})
\label{eq:projproc}
\eeq 
and the off-diagonal terms of the density matrix disappear. The 
occupation probabilities $P_n(t)$ change discontinuously at times 
$NT$ and their evolution is governed by the master equation 
\andy{eq:master}
\beq
P_n(N)=\sum_m W_{nm}P_m(N-1),
\label{eq:master}
\eeq
where we defined, with a little abuse of notation, 
\andy{defP}
\beq
P_n(N) \equiv P_n(NT+\tau)
\label{defP}
\eeq
and where
\andy{eq:transprob}
\beq
W_{nm} \equiv
|\langle n|U_{\rm free}(\tau)U_{\rm kick}U_{\rm free}(T-\tau)|m \rangle|^2=|\langle n|U_{\rm 
kick}|m \rangle|^2
\label{eq:transprob}
\eeq
are the transition probabilities.
Althought the map (\ref{eq:master})
depends on $\lambda, V, H_0$ in a complicated way, very 
general conclusions
can be drawn about the average value of a generic regular 
function of momentum $g(p)$. Let
\andy{eq:energy}
\beq
\mean{g(p)}_t \equiv {\rm Tr}(g(p)\rho(t))=
\sum_n g(p_n) P_n(t),
\label{eq:energy}
\eeq
where $p |n\rangle = p_n |n \rangle$ $(p_n=n\hbar)$,
and consider
\andy{eq:energy1}
\beq
\mean{g(p)}_{N} 
=\sum_n g(p_n) P_n(N)=\sum_{n,m} g(p_n) W_{nm}P_m(N-1),
\label{eq:energy1}
\eeq
where 
$\mean{g(p)}_N\equiv \mean{g(p)}_{NT+\tau}$
is the average value of $g$ after $N$ kicks.
Substitute $W_{nm}$ from (\ref{eq:transprob}) to obtain
\andy{eq:energy2}
\barr
\mean{g(p)}_{N} &=& 
\sum_{n,m} g(p_n) \langle m|U_{\rm kick}^\dagger|n\rangle 
\langle n|U_{\rm kick}|m\rangle P_m(N-1)\nonumber \\
&=& \sum_m \langle m|U_{\rm kick}^\dagger g(p) U_{\rm kick}|m\rangle P_m(N-1),
\label{eq:energy2}
\earr
where we used $g(p)|n\rangle=g(p_n)|n\rangle$. 
We are mostly interested in the evolution of the quantities $p$ and $p^2$.
By the Baker-Hausdorff lemma  
\andy{eq:baklemma}
\beq
U_{\rm kick}^\dagger g(p) U_{\rm kick}=g(p)+i\frac{\lambda}{\hbar}[V,g(p)]+\frac{1}{2!}\left(
\frac{i\lambda}{\hbar}\right)^2[V,[V,g(p)]]+...,
\label{eq:baklemma}
\eeq
we obtain the exact expressions \cite{note1}
\andy{eq:dhzero,dhzeroo}
\barr
U_{\rm kick}^\dagger p U_{\rm kick} &=& p+i\frac{\lambda}{\hbar}[V,p],
\label{eq:dhzero} \\
U_{\rm kick}^\dagger p^2 U_{\rm kick} &=& p^2+i\frac{\lambda}{\hbar}[V,p^2]+
\lambda^2\left(V'\right)^2,
\label{eq:dhzeroo}
\earr
where prime denotes derivative.
Substituting into (\ref{eq:energy2}) and iterating, one gets
\andy{eq:dhzero0,1}
\barr
\mean{p}_N&=&\mean{p}_{N-1}=\mean{p}_{0},\label{eq:dhzero0} \\
\mean{p^2}_N&=&\mean{p^2}_{N-1} +\lambda^2 \mean{f^2}
=\mean{p^2}_{0} +\lambda^2 \mean{f^2}N,
\label{eq:dhzero1}
\earr
where $f=-V'(x)$ is the force and
\andy{meanforce}
\beq
\mean{f^2}= {\rm Tr}\left(f^2\rho_{NT+\tau}\right)=
\sum_n\langle n|f^2|n\rangle P_n(N)=\frac{1}{2\pi}\int_{-\pi}^\pi dx\;f^2(x)
\label{eq:meanforce}
\eeq
is a constant that does not depend on $N$ because $\langle n|f^2|n\rangle$
is independent of the state $|n\rangle$ [see (\ref{eq:eigenfun})] 
and $\sum P_n=1$.
In particular, the kynetic energy $K=p^2/2m$
grows at a constant rate:
$\mean{K}_{N}=\mean{K}_{0}+\lambda^2\mean{f^2} N/2m$.
By using (\ref{eq:dhzero0})-(\ref{eq:dhzero1}) we obtain the   
friction coefficient
\andy{drift}
\beq
F=\frac{\mean{p}_N-\mean{p}_0}{NT}=0
\label{eq:drift}
\eeq
and the diffusion coefficient
\andy{diffusion}
\beq
D=\frac{\mean{\Delta p^2}_N-\mean{\Delta p^2}_0}{NT}
=\frac{\lambda^2 \mean{f^2}}{T},
\label{eq:diffusion}
\eeq
where $\mean{\Delta p^2}_N=\mean{p^2}_N-\mean{p}_N^2$. The above 
results are {\em exact}: their derivation involves no 
approximation. This shows that this class of Hamiltonian systems, 
if ``measured" after every kick, has a constant diffusion rate in 
momentum with no friction, for {\em any} perturbation $V=V(x)$. 

In particular, in the seminal kicked-rotator model,
one gets ($H_0=p^2/2I$ and $V=\cos x$)
\andy{eq:krdiff}
\beq
D= \frac{\lambda^2}{2T} :
\label{eq:krdiff}
\eeq
this is nothing but the diffusion constant obtained in the 
classical case \cite{Chirikov,KaulGontis}. Notice that one obtains 
the quasilinear diffusion constant without higher-order correction 
\cite{Lichten}. 

The above results may seem somewhat puzzling, essentially because 
one finds that in the quantum case, when repeated measurements of 
momentum (action variable) are performed on the system, a chaotic 
behavior is obtained for {\em every} value of $\lambda$ and for 
{\em any} potential $V(x)$. On the other hand, in the classical 
case, diffusion occurs only for some $V(x)$, when $\lambda$ exceeds 
some critical value $\lambda_{\rm crit}$. (For instance, the kicked 
rotator  displays diffusion for $\lambda \geq \lambda_{\rm 
crit}\simeq 0.972$ \cite{Chirikov,Lichten}.) It appears, therefore, 
that quantum measurements not only yield a chaotic behavior in a 
quantum context, they even produce chaos when the classical motion 
is regular. In order to bring to light the causes of this peculiar 
situation, it is necessary to look at the classical case. The 
classical map for the Hamiltonian (\ref{eq:hamilt}) reads 
\andy{eq:classmap}
\barr
x_{N} &=& x_{N-1}+H'_0(p_{N-1})T, \nonumber \\
p_{N} &=& p_{N-1} -\lambda V'(x_{N}).
\label{eq:classmap}
\earr

A quantum measurement of $p$ yields an exact determination of 
momentum $p$ and, as a consequence, makes position $x$ completely 
undetermined (uncertainty principle). This situation has no 
classical analog: it is inherently quantal. However, the classical 
``map" that best mymics this physical picture is obtained by 
assuming that position $x_N$ at time $\tau$ after each kick (i.e.\ 
when the quantum counterpart undergoes a measurement) behaves like 
a random variable $\xi_N$ uniformly distributed over $[-\pi,\pi]$: 
\andy{eq:classmeas}
\barr
x_{N}&=&\xi_{N}, \nonumber \\ p_{N}&=&p_{N-1}-\lambda V'(x_{N}). 
\label{eq:classmeas}
\earr
Introducing the ensemble average $\langle\langle 
\cdots\rangle\rangle$ over the stochastic process
(i.e.\ over the set of independent random variables $\{\xi_k 
\}_{k\leq N}$), it is straightforward to obtain 
\andy{eq:meanenergy}
\barr
\langle\langle p_N \rangle\rangle &=&
\langle\langle p_{N-1} \rangle\rangle
-\lambda \langle V'(\xi_{N}) \rangle,\nonumber\\
\langle\langle \Delta p_N^2\rangle\rangle &=&
\langle\langle\Delta p_{N-1}^2\rangle\rangle
+\lambda^2\left(\langle V'(\xi_{N})^2\rangle-\langle V'(\xi_{N})\rangle^2\right),
\label{eq:meanenergy}
\earr
where $\Delta p^2_N=p^2_N-\langle\langle p_N\rangle\rangle^2$ and
\andy{eq:meandef}
\beq
\langle g(\xi) \rangle \equiv
\frac{1}{2\pi}\int_{-\pi}^\pi g(\xi) d\xi
\label{eq:meandef}
\eeq
is the average over the single random variable $\xi$ 
[this coincides with the quantum average: see for instance the
last term of (\ref{eq:meanforce})].
In deriving (\ref{eq:meanenergy}),
the average of $V'(\xi_N) p_{N-1}$ was factorized because $p_{N-1}$ 
depends only on $\{\xi_k \}_{k\leq N-1}$. 
The average of $V'(\xi_N)$ 
in (\ref{eq:meanenergy}) vanishes due to the periodic boundary conditions 
on $V$, so that 
\andy{eq:mean22}
\beq
\langle\langle \Delta p_N^2\rangle\rangle =
\langle\langle\Delta p_{N-1}^2\rangle\rangle
+\lambda^2 \langle f^2\rangle
\label{eq:mean22}
\eeq
and the momentum diffuses at the rate (\ref{eq:diffusion}), as in 
the quantum case with measurements. What we obtain in this case is 
a diffusion taking place in the whole phase space, without effects 
due to the presence of adiabatic islands. 

It is interesting to frame our conclusion in a proper context, by 
comparing the different cases analyzed: (A) a classical system, 
under the action of a suitable kicked perturbation, displays a 
diffusive behavior if the coupling constant exceeds a certain 
threshold (KAM theorem); (B) on the other hand, in its quantum 
counterpart, this diffusion is always suppressed. (C) The 
introduction of measurements between kicks encompasses this 
limitation, yielding diffusion in the quantum case. More so, 
diffusion takes place for any potential and all values of the 
coupling constant (namely, even when the classical motion is 
regular). (D) The same behavior is displayed by a ``randomized 
classical map," in the sense explained above. These conclusions are 
sketched in Table~1. 
\begin{center}
{\small {\bf Table 1}: Classical vs quantum diffusion}  \\  \quad \\
\begin{tabular}{|c|c|c|}    
  \hline\hline 
A &classical         & diffusion for $\lambda > \lambda_{\rm crit}$ \\
  \hline
B  &quantum           & no diffusion  \\
  \hline
C  &quantum + measurements & diffusion $\forall \lambda$ \\
  \hline
D  &classical + random     & diffusion $\forall \lambda$ \\
  \hline\hline
\end{tabular} 
\end{center}
As we have seen, the effect of measurements is basically equivalent 
to a complete randomization of the classical angle variable $x$, at 
least for the calculation of the diffusion coefficient in the 
chaotic regime. There are two points which deserve clarification. 
Indeed, one might think that: i) the  randomized classical map 
(\ref{eq:classmeas}) and the quantum map with measurements 
(\ref{eq:master}), (\ref{eq:dhzero0})-(\ref{eq:diffusion}) are 
identical; ii) the diffusive features in a quantum context are to 
be ascribed to the projection process (\ref{eq:projproc}) (hence to 
a non-unitary dynamics). Both expectations would be incorrect. As 
for i), there are corrections in $\hbar$: it is indeed 
straightforward to show that the two maps have equal moments up to 
third order, while the fourth moment displays a difference of order 
$O(\hbar^2)$: 
\andy{fourth}
\beq
\langle p^4 \rangle_N - \langle p^4 \rangle_{N-1}
=\langle\langle p_N^4\rangle\rangle
-\langle\langle p_{N-1}^4\rangle\rangle
+\lambda^2\hbar^2\langle (f')^2\rangle.
\label{eq:fourth}
\eeq
As for ii), it suffices to observe that the very same results can 
be obtained by making only use of a purely unitary evolution. To 
this end, we must give a model for measurement, by looking more 
closely at the physics of such a process. When a quantum 
measurement is performed, the relevant information is recorded in 
an apparatus. For example, the measured system scatters one or more 
photons   (phonons) and each $p$-eigenstate gets entangled with the 
photon (phonon) wave function. A process of this sort can be 
schematized by associating an additional degree of freedom (a 
``spin" is the simplest possible case) with every momentum 
eigenstate, at time $\tau$ after every kick. This is easily 
accomplished by adding the following ``decomposition" \cite{Wigner} 
Hamiltonian to (\ref{eq:hamilt}) 
\andy{eq:decham}
\beq
H_{\rm dec} = \frac{\pi}{2} \sum_{n,k} |n\rangle\langle n|
\otimes \sigma^{(n,k)} \delta (t-kT-\tau),
\label{eq:decham}
\eeq
where $|n\rangle$ is an eigenstate of $p$ and $\sigma^{(n,k)} \;\;
\forall (n,k)$ is the first Pauli matrix, whose action is given by
\andy{eq:sigma1}
\beq
\sigma^{(n,k)} |\pm \rangle_{(n,k)} = |\mp \rangle_{(n,k)},
\label{eq:sigma1}
\eeq
where $|+ \rangle_{(n,k)}, |- \rangle_{(n,k)}$ denote spin up, 
down, respectively, in ``channel" $(n,k)$. Let us prepare the 
system in the initial ($t=0^+$) state 
\andy{eq:inst}
\beq
|\Psi_{\rm in} \rangle = \sum_m c_m |m \rangle \bigotimes_{k,n} |- 
\rangle_{(n,k)} 
\label{eq:inst}
\eeq
(all ``spins" down). For the sake of simplicity, we shall  
concentrate  our attention on the first two kicks. In the same 
notation as in (\ref{defP}), the evolution of the state $|\Psi (N) 
\rangle \equiv |\Psi (NT+ \tau^+) \rangle$ reads 
\andy{eq:inst1}
\beq
|\Psi (0) \rangle
= -i \sum_m c^\prime_m |m \rangle \otimes |+\rangle_{(m,0)} 
\bigotimes_{k\geq 1,n} |- \rangle_{(n,k)} \ ,
\label{eq:inst1}
\eeq
\andy{eq:inst2}
\beq
|\Psi (1) \rangle  
= (-i)^2 \sum_{\ell,m} |\ell \rangle 
\otimes |+\rangle_{(\ell,1)} \otimes A_{\ell m}
c^\prime_m |+\rangle_{(m,0)} 
\bigotimes_{k\geq 2,n} |- \rangle_{(n,k)} \ ,
\label{eq:inst2}
\eeq
\andy{eq:inst3}
\beq
|\Psi (2) \rangle 
= (-i)^3 \sum_{j,\ell,m} |j \rangle 
\otimes |+\rangle_{(j,2)}\otimes A_{j \ell}|+\rangle_{(\ell,1)}
\otimes A_{\ell m} c^\prime_m |+\rangle_{(m,0)} 
\bigotimes_{k\geq 3,n} |- \rangle_{(n,k)} \ ,
\label{eq:inst3}
\eeq
where $c^\prime_m = c_m \exp [-i H_0(p_m) \tau ]$ and 
\andy{eq:transampl}
\beq
A_{\ell m} \equiv \langle \ell|U_{\rm free}(\tau)U_{\rm kick}U_{\rm 
free}(T-\tau)|m \rangle 
\label{eq:transampl}
\eeq
is the transition amplitude ($W_{\ell m} = |A_{\ell m}|^2$). We see 
that at time $\tau$ after the $k$th kick, the $n$th eigenstate of 
the system becomes associated with spin up in channel $(n,k)$. By 
using (\ref{eq:inst2})-(\ref{eq:inst3}) one readily shows that the 
occupation probabilities evolve according to  
\andy{eq:transprobev}
\beq
P_n(2) \equiv \langle \Psi(2) | \Big(|n \rangle\langle n| \otimes 
{\bf 1}_{\rm spins} \Big) |\Psi(2)\rangle = \sum_m W_{nm} P_m(1). 
\label{eq:transprobev}
\eeq
The generalization to $N$ kicks is straightforward and it is very 
easy to obtain the same master equation (\ref{eq:master}). The 
observables of the quantum particle  evolve therefore like in 
(\ref{eq:energy1}): in particular, the average value of the quantum 
observable $\tilde p = p \otimes {\bf 1}_{\rm spins}$ displays 
diffusion with coefficients (\ref{eq:drift})-(\ref{eq:diffusion}). 
This shows that projection operators are not necessary to obtain a 
quantal diffusive behavior and the {\em unitary} dynamics 
engendered by (\ref{eq:hamilt}) and (\ref{eq:decham}) yields the 
same results. 

Our analysis can be easily generalized to radial twisting maps in
higher dimensions. It would be interesting to extend it to a slightly
different class of Hamiltonians, such as those 
used in \cite{Casati} to analyze the effect of
an oscillating perturbation on an atomic system.

\vspace*{.5cm}
We thank Hiromichi Nakazato and Mikio Namiki for early discussions.


\begin{thebibliography}{00}
\bibitem{note1} \andy{note1}
In general, for  polynomial $g(p)$, the highest order of $\lambda$ 
appearing in (\ref{eq:baklemma}) is the degree of the polynomium.
\end{thebibliography}
\end{document}